\documentclass[10pt]{IEEEtran}
\usepackage{algorithm}
\usepackage{algpseudocode}
\usepackage{cite}
\usepackage{bm}
\usepackage{amsmath}
\usepackage{extarrows}
\usepackage{amssymb}
\usepackage{graphicx}
\usepackage{color}
\usepackage{bookmark}
\graphicspath{{figure/}}

\newcommand{\mr}{\mathrm}

\newcommand{\BE}{\begin{equation}}
\newcommand{\EE}{\end{equation}}
\newcommand{\BS}{\begin{subequations}}
\newcommand{\ES}{\end{subequations}}
\renewcommand{\bf}{\bm}

\newtheorem{proposition}{Proposition}

\newtheorem{definition}{Definition}

\newtheorem{lemma}{Lemma}

\makeatletter

\newcommand{\Rmnum}[1]{\expandafter\@slowromancap\romannumeral #1@}
\makeatother

\ifCLASSINFOpdf

\else

\fi

\begin{document}

\title{A New Insight into GAMP and AMP
}

\author{\IEEEauthorblockN{ Lei Liu, \emph{Member, IEEE}, Ying Li, \emph{Member, IEEE}, Chongwen Huang, \emph{Student Member, IEEE},\\ Chau Yuen, \emph{Senior Member, IEEE,} Yong Liang Guan, \emph{Senior Member, IEEE} }
\thanks{Lei Liu is with the Department of Electronic Engineering, City University of Hong Kong, Hong Kong (e-mail: leiliu@cityu.edu.hk). }
\thanks{Ying Li is with the State Key Lab of Integrated Services Networks, Xidian University, Xi'an 710071, China (e-mail: yli@mail.xidian.edu.cn).}
\thanks{Chongwen Huang and Chau Yuen are with the Singapore University of Technology and Design, Singapore 487372 (e-mail: chongwen\_huang@mymail.sutd.edu.sg; yuenchau@sutd.edu.sg).}
\thanks{Yong Liang Guan is with the School of Electrical and Electronic Engineering, Nanyang Technological University, Singapore 639798 (e-mail: eylguan@ntu.edu.sg).}}

\maketitle
\begin{abstract}
A concise expectation propagation (EP) based message passing algorithm (MPA) is derived for the general measurement channel. By neglecting some high-order infinitesimal terms, the EP-MPA is proven to be equivalent to the Generalized Approximate Message Passing (GAMP), which exploits central limit theorem and Taylor expansion to simplify the belief propagation process. Furthermore, for additive white gaussian noise measurement channels, EP-MPA is proven to be equivalent to the AMP. Such intrinsic equivalence between EP and GAMP/AMP offers a new insight into GAMP and AMP via a unified message passing rule for non-linear processing, and may provide clues towards building new MPAs in solving more general non-linear problems.
\end{abstract}


\IEEEpeerreviewmaketitle
\section{Introduction}
Generalized approximate message passing (GAMP) proposed by Rangan \cite{Rangan2010,Rangan2011} is a generalization of approximate message passing (AMP), independently described by Donoho et al. \cite{Donoho2009}. The GAMP allows general measurement channels (including non-linear channels) to be used. Due to its Bayes optimality as well as low computational complexity, and more importantly, asymptotical accuracy of state evolution (SE), GAMP has attracted more and more attention in domains like compressive sensing, image processing, Bayesian learning, statistical physics, low-rank matrix estimation, mmWave channel estimation, spatial modulation, user activity and signal detection in random access, orthogonal frequency division multiplexing  analog-to-digital converters system, sparse superposition codes, etc \cite{Eliasi2017, Fang2016, Metzler2015, Yuwei2018, Biyik2017, Zhao2018, Kokshoorn2018, Cao2017}.

The original AMP and GAMP are derived via belief propagation (BP) based on the central limit theorem (CLT) and Taylor Series. Expectation propagation (EP) \cite{Minka2001,opper2005expectation} is an alternative message passing rule that deals with general {non-Gaussian probability distribution functions (PDFs)}. EP projects the \emph{a-posteriori} estimation on a Gaussian distribution with moment matching, and thus obtains a similar message update rule as Gaussian message passing (GMP)\cite{Loeliger2006, Lei2016a, Lei2016b,Lei20161b, Yuhao2018}. The potential connection between AMP and EP was first shown in \cite{Cakmak2014, Heskes2005}, in which the fixed points of EP and AMP were shown to be consistent. An EP-based AMP was proposed in \cite{Kuang2014}. Recently, Ma and Ping proposed an orthogonal AMP for general unitarily-invariant measurement matrices, and showed that the optimal MMSE OAMP is equivalent to MMSE EP \cite{Ma2017, Takeuchi2017, Rangan2016}. These works hint at the conceptual equivalence between EP and AMP. In \cite{Meng2015}, Meng et al. first gave a rigorous derivation of AMP based on a dense graph-based EP by making some approximations in large system limit. Based on the results in \cite{Meng2015}, the authors further provided a unified Bayesian inference framework for the extension of AMP and VAMP to the generalized linear model \cite{Meng2018SPL, Meng2018access}.
Another form of EP-based derivation for MMSE GAMP was illustrated in \cite{QZou2018}. More recently, the connection between EP and the max-sum GAMP was built in \cite{Zhu2019}. 

In \cite{Rangan2010,Rangan2011,Donoho2009}, the authors used Taylor expansion and second-order approximation for the non-linear constraints of the general measurement channel. In this paper, we adopt a different approach, in which the general non-linear constraints are solved by an easily understandable EP rule, which has the same form as the GMP rule (for the linear constraints). The only difference between EP and GMP is that the \emph{a-posteriori} calculation is replaced by a non-linear MMSE estimation, which makes EP more efficient in solving the non-linear problems than GMP. As a result, the whole general measurement problem is solved by the unified ``GMP-like" rule. By neglecting the high-order infinitesimal terms, the EP-MPA is proven to be equivalent to GAMP. Furthermore, for additive white Gaussian noise (AWGN) measurement channels, the EP-MPA is proven to be equivalent to AMP. These results offer a new insight into GAMP and AMP, and may provide hints to build new MPAs for more general non-linear networks.

We credit \cite{Meng2015, Meng2018SPL, Meng2018access, QZou2018,Zhu2019} for the work on the consistency between EP and GAMP/AMP. However, this correspondence firstly provides a unified ``GMP-like" rule for the MPAs in solving general measurement channels.


\emph{Notations:} Let $a_{mn}$ denote the $(m,n)$-th entry of matrix $\bf{A}$, $a_i$ the $i$-th entry of vector $\bf{a}$, $\langle\cdot\rangle$ the average value operation, $(\cdot)^H$ conjugate transpose, $\lim\limits_{n\to 0 } {\mathcal{O}(n)}/{n}\to \mathrm{constant}$, $\lim\limits_{n\to 0 }{o(n)}/{n}\to 0$, and $\mr{E}\{a|b\}$ and $\mr{var}\{a|b\}$ the conditional expectation and variance.

\section{Problem Formulation}
\begin{figure}[htb]
  \centering
  \includegraphics[width=8cm]{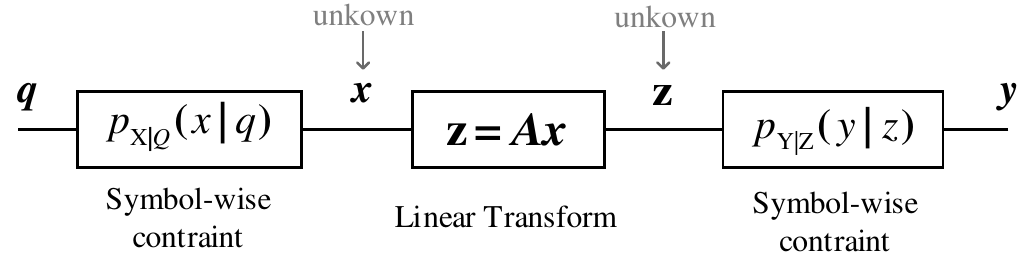}\\
  \caption{System model: $\bf{x}$ and $\bf{z}$ are subjected to an linear function $\bf{z}=\bf{A}\bf{x}$, and $\bf{x}$ and $\bf{z}$ are subjected to  symbol-wise transfer probability functions.}\label{diag_graph}
\end{figure}
GAMP considers a system given in Fig. \ref{diag_graph}, where $\bf{x}\in\mathbb{R}^N$, $\bf{A}\in\mathbb{R}^{M\times N}$, and $\bf{z}\in\mathbb{R}^M$ are subjected to a linear function $\bf{z}=\bf{A}\bf{x}$, and $\bf{x}$ and $\bf{z}$ are subjected to  symbol-wise transfer probability function $p(\bf{x}|\bf{q})=\prod\limits_{n=1}^N {p_{X|Q}(x_n|q_n)}$  and  $p(\bf{y}|\bf{z})=\prod\limits_{m=1}^M {p_{Y|Z}(y_m|z_m)}$ respectively. In addition, $\bf{A}$ has i.i.d. Gaussian components $a_{ij}\sim \mathcal{CN}(1,1/M)$. The goal of GAMP is to iteratively recover $\bf{x}$ and $\bf{z}$  given $\bf{q}$ and $\bf{y}$, which is equivalent to estimate the marginal probability below\vspace{-0.2cm}
\BS\label{Eqn:post_proba}\begin{align}
&p(\bf{x}, \!\bf{z}|\bf{y},\!\bf{q}) \propto p(\bf{y}|\bf{x})p(\bf{x}|\bf{q})  =\prod\limits_{m=1}^{M} {p_{Y|Z}\left(y_m|[\bf{A}\bf{x}]_{m}\right)} \\
 &= 
\delta (\bf{A}\bf{x}\!-\!\bf{z})\!\! \prod\limits_{m=1}^{M} \!\!{p_{Y|Z}\!\left(y_m|z_{m}\right)} \!\!\prod\limits_{n=1}^{N}\!\! {p_{X|Q}\!(x_n|q_n)},
\end{align}\ES
where $\delta(\cdot)$ is a Dirac delta function. However, exact calculation of \eqref{Eqn:post_proba} has intractable complexity for large scale problems.

{For more general $a_{ij}\sim \mathcal{CN}({0},\sigma^2_a/M)$ with finite $\sigma^2_a$, we can rewrite the system to ${\bf{y}}' = {\bf{y}}/\sigma_a = {\bf{A'x}} +{\bf{n}}'= \sigma_a^{-1}{\bf{Ax}} + \sigma_a^{-1}{\bf{n}}$, where $ a'_{ij}\sim \mathcal{CN}({0},1/M)$ and $\bm{n}'\!\sim\!\mathcal{CN}(\mathbf{0},\sigma^2\sigma_a^{-2}\bm{I})$. Then, all the results in this paper are still valid by replacing $\sigma^2$ with $\sigma^2\sigma_a^{-2}$. For example, if $a_{ij}\sim \mathcal{CN}({0},1/N)$, we replace $\sigma^2$ by ${N\sigma^2}/{M}$ to make the results of this paper be valid.}

\section{EP-Based Message Passing Algorithm}\label{SEC:EP}
\begin{figure}[hbt]
  \centering
  \includegraphics[width=7cm]{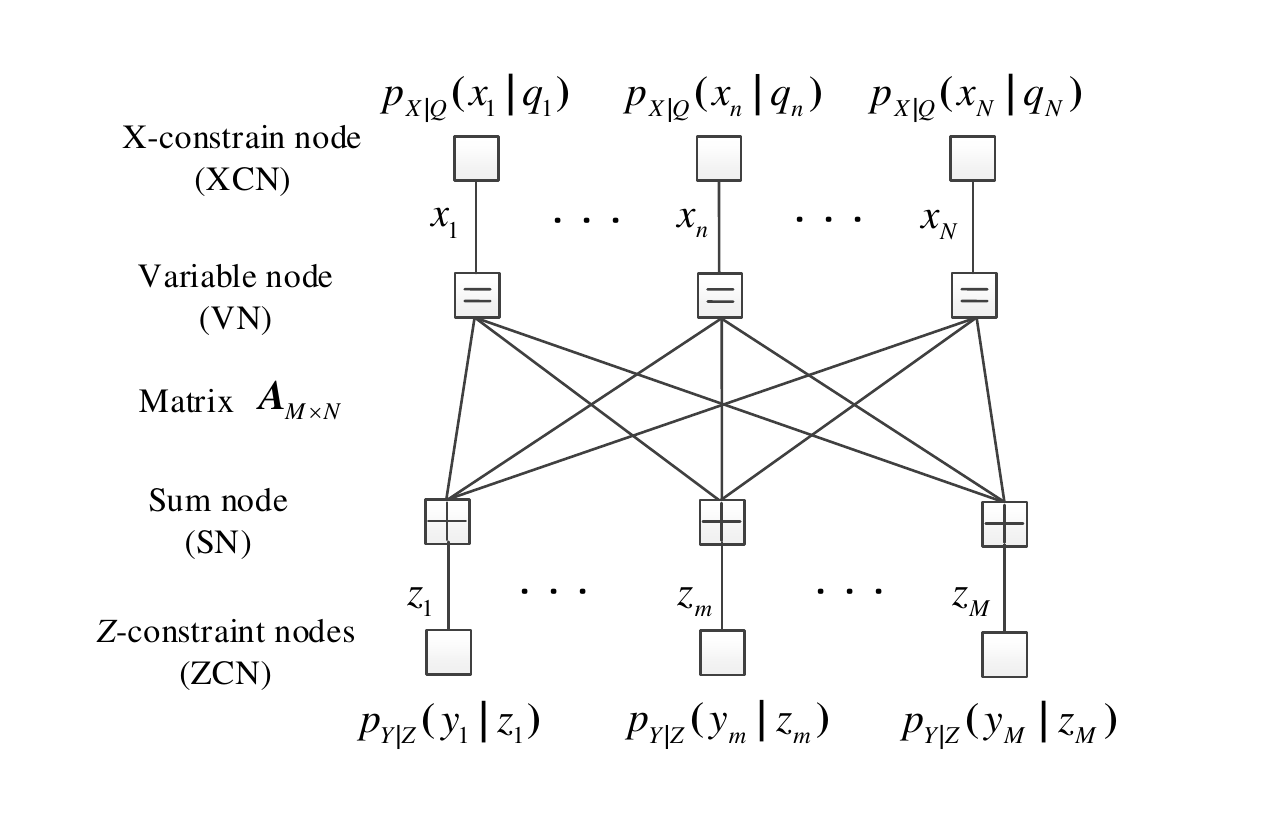}\\\vspace{-0.2cm}
  \caption{{Forney-style factor graph}. Edges denote variables, and nodes denote the related constraints: $p(x_n|q_n)$, ${p(y_m|z_m)}$ and $y_m=\sum\limits_{n=1}^N{a_{mn}x_n}$.}\label{fact_graph}
\end{figure}
{Fig. \ref{fact_graph} gives a {Forney-style factor graph} of the system in \eqref{Eqn:post_proba}, where edges denote variables, and nodes denote the related constraints: $p(x_n|q_n)$, ${p(y_m|z_m)}$, $x_{n1}=\dots=x_{nM}$ and $y_m=\sum\limits_{n=1}^N{a_{mn}x_n}$.} MPA \cite{Loeliger2006} is a method to iteratively compute the marginal probability. Since the high-dimensional integration is distributively calculated by local message passing, it has a low complexity. Next, we briefly introduce EP \cite{Minka2001,opper2005expectation}.

\subsection{Expectation Propagation}
\begin{definition}
Let the \emph{a-priori} message be ${x}_{\mr{in}}={x}+v_{\mr in}^{1/2}{w}$ with ${w}\sim\mathcal{CN}({0},1)$, and ${x}\in \mathcal{X}$ a constraint of $x$. EP updates\vspace{-0.2cm}
\BS\label{Eqn:EP}\begin{align}
 {x}_{\mr{out}}  &=\!v_{\mr{out}} \left[ {v_{\mr{post}}^{-1}} {x_{\mr{post}}}  \!-\! {v_{\mr{in}}^{-1}}{{x}_{\mr in}}\right], \\
v_{\mr{out}} &=\!\left[{v_{\mr{post}}^{-1}} \!-\! {v_{\mr{in}}^{-1}}\right]^{-1}\!\!,
\end{align}\ES
where $x_{\mr{post}}\equiv{\mr{E}}\{{x}|{x}_{\mr{in}}, \mathcal{X}\}$ and ${{v_{\mr{post}}}\equiv \mr{var}\{x|x_{\mr{in}},\mathcal{X}\} }$.
\end{definition}

{By letting $v_i=v_{\mr{out}}$, $m_i= x_{\mr{out}}$, $v_{\theta}=v_{\mr{in}}$, $m_{\theta}= x_{\mr{in}}$, $v_{\theta}^{new}=v_{\mr{post}}$ and $m_{\theta}^{new}= x_{\mr{post}}$, it is easy to verify that \eqref{Eqn:EP} is consistent with that in \cite{Minka2001} (see Eqs. 3.32-3.34). The form in \eqref{Eqn:EP} has also been widely used for EP \cite{MaISTC, Takeuchi2017}.}

{\underline{\emph{\textbf{Relation to Standard GMP:}}} In fact, when the constraint ${x}\in \mathcal{X}$ is a linear and Gaussian\footnote{For example, $\mathcal{X}=\{x|x\in \mathcal{CN}(m_x,v_x)\}$ is a Gaussian constraint of $x$, and $\mathcal{X}=\{x|y=ax+ b\}$ (given $y$, $a$ and the distribution of $b$) is a linear constraint of $x$.}, EP in \eqref{Eqn:EP} is the exact GMP. For example, if $\mathcal{X}$ is a Gaussian constraint ${x}\sim \mathcal{CN}(m_x, v_x)$, the \emph{a posteriori probability} is Gaussian and given by
\BS\begin{align}
p(x|x_{\mr{in}},{x}\in \mathcal{X}) &\propto e^{-\frac{|x-m_x|^2}{v_x}} e^{-\frac{|x-{x}_{\mr{in}}|^2}{{v}_{\mr{in}}}}\\
&\propto e^{-[v_x^{-1}+v_{\mr{in}}^{-1}]|x|^2 + 2 [v_x^{-1}m_x+v_{\mr{in}}^{-1}x_{\mr{in}}]x} \\
&\propto  e^{-\frac{|x-x_{\mr{post}}|^2}{v_{\mr{post}}}}
\end{align}\ES}\vspace{-0.2cm}
{where\vspace{-0.2cm}
\BS\label{Eqn:GMP}\begin{align}
 {x}_{\mr{post}}  &=\!v_{\mr{post}} \left[ {v_{x}^{-1}} {m_{x}} + {v_{\mr{in}}^{-1}}{{x}_{\mr in}}\right], \\
v_{\mr{post}} &=\!\left[{v_{x}^{-1}} + {v_{\mr{in}}^{-1}}\right]^{-1}\!\!,
\end{align}\ES
which can be rewritten to\vspace{-0.2cm}
\BS \begin{align}
  m_x & =  v_{\mr{out}} \left[ {v_{\mr{post}}^{-1}} {x_{\mr{post}}}  \!-\! {v_{\mr{in}}^{-1}}{{x}_{\mr in}}\right], \\
  v_x & = \left[{v_{\mr{post}}^{-1}} \!-\! {v_{\mr{in}}^{-1}}\right]^{-1}\!\!.
\end{align}\ES
GMP \cite{Loeliger2006, Lei2016a, Lei2016b,Lei20161b} follows the well-known extrinsic message passing (EMP), named Turbo principle, where the output does not involve the input $[x_{\mr{in}},v_{\mr{in}}]$, i.e.,
\BS\label{Eqn:WP}\begin{align}
{x}_{\mr out}  &= m_x = {{\mr{E}}\{{x}| {x}\in \mathcal{X}\}}, \\
v_{\mr{out}} &= v_x = {{\mr{var}}\{{x}| {x}\in \mathcal{X}\}}.
\end{align}\ES
From \eqref{Eqn:GMP}, \eqref{Eqn:WP} is the same as \eqref{Eqn:EP}. Hence, GMP is an instance of EP. In Turbo, there is a famous ``information equation":\vspace{-0.2cm}
\BE\label{Eqn:inf_eq}
``\mathrm{Extrinsic}" = ``\mathrm{Post}" - ``\mathrm{Priori}".\vspace{-0.2cm}
\EE
{That is, the information contained in the \emph{a-posteriori} message is equal to the sum information contained in the \emph{a-priori} message and the extrinsic message. This principle has been widely used in modern channel coding and sum-product algorithm. For example, the extrinsic message can be calculated by removing the \emph{a-priori} message from the \emph{a-posteriori} message.}

If ${x}\in \mathcal{X}$ is non-Gaussian, EP in \eqref{Eqn:EP} is not equal to GMP, i.e., \eqref{Eqn:EP} and \eqref{Eqn:WP} are not equivalent, i.e., ``information equation" in \eqref{Eqn:inf_eq} does not hold any more. In general, EP could provide more useful information than EMP (or Turbo) for non-Gaussian $\mathcal{X}$, i.e., the following ``information inequality" holds:\vspace{-0.2cm}
\BE
``\mathrm{Post}" - ``\mathrm{Priori}" > ``\mathrm{Extrinsic}",\vspace{-0.2cm}
\EE
which implies that ``EP" outperforms ``Turbo". For more details, refer to \cite{MaISTC, Lei_Capacity}.}

\underline{\emph{\textbf{Intuition of EP:}}} In general, the {\emph{a posteriori probability}  (APP) estimation}  is the optimal local estimation since it fully exploits the \emph{a-priori} (or input) message, but it will cause correlation problem in the iterative process. To avoid the correlation problem in the iteration, Turbo principle discards the \emph{a-priori} message in the estimation, but this results in performance loss since the  \emph{a-priori} message is not exploited. EP makes a good tradeoff between the APP and Turbo, i.e., the \emph{a-priori} message is partly used to improve the estimation and the correlation problem is also avoided. {Due to these reasons, EP could have a better performance than EMP.}

{Fig. \ref{Fig:Message_passing} shows the message passing illustration for the problem, where $({x}_{n}^v, v_{n}^v)$ be the messages (mean and variance for $x_n$) passing from VN to XCN, $(\tilde{x}_{n}, \tilde{v}_{n}^x)$ for $x_n$ from XCN to VN, $(x_{mn}^v, v_{mn}^v)$ for $x_n$ from VN to SN, $(x_{mn}^s, v_{mn}^s)$ for $x_n$ from SN to VN, and $({z}_{m}^s, v_{m}^s)$ for $z_m$ from SN to ZCN, and $(\tilde{z}_{m}, \tilde{v}_{m}^z)$ for $z_m$ from ZCN to SN. Next, we derive the message passing algorithm based on the expectation propagation principle under a unified ``GMP-like" rule.}

\subsection{A Unified ``GMP-like" EP-MPA}
\begin{figure}[t]
  \centering
  \includegraphics[width=8.5cm]{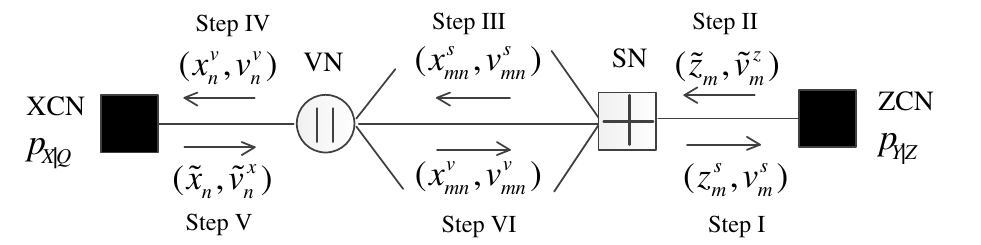}\\
  \caption{EP-MPA illustration.}\label{Fig:Message_passing}
\end{figure}

Fig. \ref{Fig:Message_passing} illustrates the EP-MPA, where ZCN and XCN denote the constraint nodes of $z$ and $x$ respectively. The message updates at variable node (VN) and sum node (SN) are GMP, while ZCN and XCN are EP.

\underline{\emph{\textbf{Step I (SN $\to$ ZCN):}}}
Since $z_m=\sum_{n}{a_{mn}{x}_{n}}$ and $x^v_{mn}(t)=x_n+v^v_{mn}(t)^{1/2}w$, from central limit theorem (CLT), we have $z^s_{m}(t)= z_m+v^s_{m}(t)^{1/2}w$, where
\BS\begin{align}
z^s_{m}(t)=\sum_{n}{a_{mn}{x}^v_{mn}}(t), \;\;
v^s_{m}(t)=\sum_{n}{|a_{mn}|^2{v}^v_{mn}}(t),
\end{align}\ES
with initialization ${x}^v_{mn}(1)\!=\!{\rm E}\{x_n|q_n\}$, ${v}^v_{mn}(1)\!=\!{\rm{var}}\{x_n|q_n\}$.

\underline{\emph{\textbf{Step II (ZCN $\to$ SN):}}}
Message update at ZCN for $\bf{z}$ uses EP with constraints $z^s_{m}(t)= z_m+v^s_{m}(t)^{1/2}w$ and $p(y_m|z_m)$:
\BS\begin{align}\label{Eqn:ZCN2SN}
\!\!\!\!\tilde{z}_{m}(t) &\!=\! \tilde{v}^z_m(t)  \left[\frac{{\rm E}\{z_m|z_m^s(t),v_m^s(t);y_m\}}{{\rm var}\{z_m|z_m^s(t),v_m^s(t);y_m\}} \!-\! \frac{z_m^s(t)}{v_m^s(t)}\right],\\
\!\!\!\!\tilde{v}_m^z(t) &\!=\! \left[ {\rm var}\{z_m|z_m^s(t),v_m^s(t);y_m\}^{-1}\! -\! ({v_m^s}(t))^{-1} \right]^{-1}\!\!,
\end{align}\ES
where $\tilde{z}_{m}(t)\approx z_m+\tilde{v}_m^z(t)^{1/2}w$.

\underline{\emph{\textbf{Step III (SN $\to$ VN):}}}
The constraints at $m$-th SN are $z_m=\sum_{n}{a_{mn}{x}_{n}}$, $\tilde{z}_{m}(t)\approx z_m+\tilde{v}_m^z(t)^{1/2}w$ and $x^v_{mn}(t)=x_n+v^v_{mn}(t)^{1/2}w,   \forall n$. Message update at SN for VN are:
\BS\label{Eqn:SN2VN}\begin{align}
x^s_{mn}(t)  &  = { a_{mn}^{-1}\big(\tilde{z}_{m}(t)- z_{m}^s(t) +a_{mn} x_{mn}^v(t)\big)},   \\
v^s_{mn}(t) & \approx { |a_{mn}|^{-2}\big( \tilde{v}^z_{m}(t) + v_{m}^s(t) \big)},\label{Eqn:SN2VN2}
\end{align}\ES
where  ${v}^v_{mn}(t)\!\ll\! v_{m}^s(t)$ and $x^s_{mn}(t)\!=\! x_n\!+\!{v}_{mn}^s(t)^{1/2}w$.

\underline{\emph{\textbf{Step IV (VN $\to$ XCN):}}} 
The constraints at $n$-th VN are $x^s_{mn}(t)= x_n+{v}_{mn}^s(t)^{1/2}w ,\forall m$. Message update at VN are:
\BS\label{Eqn:VN2XCN}\begin{align}
x_{n}^v(t) &=\! v_{n}^v (t){\sum}_{m} { \frac{x_{mn}^{s}(t)} {v^s_{mn}(t)} }, \\
 v_{n}^v (t)&=\! \Big({\sum}_{m} \frac{1} {v^s_{mn}(t)} \Big)^{-1},
\end{align}\ES
where $x_{n}^v(t) = x_n+ v_{n}^v(t)^{1/2}w$.

\begin{algorithm}[b!]
\caption{A unified ``GMP-like" EP-MPA}\label{Alg:EP}
\begin{algorithmic}[1]
\State {\small{\textbf{Input:}   $\epsilon>0$, $N_{ite}^{ese}$}}, $\mathbf{A}$, $\bf{y}$, $\{p(x_n|q_n)\}, \{{p(y_m|z_m)}\}$.
\State \textbf{Initialization:}  $t=1$, $\{{x}^v_{mn}(1)={\rm E}\{x_n|q_n\}$, ${v}^v_{mn}(1)={\rm{var}}\{x_n|q_n\}, \forall n,\forall m\}$.
\State \textbf{Do}
\State \quad \underline{\emph{\textbf{Step I}}}: For each $m$ compute: \[\begin{array}{l} v^s_{m}(t)={\sum}_{n}{|a_{mn}|^2{v}^v_{mn}}(t), \;
z^s_{m}(t)={\sum}_{n}{a_{mn}{x}^v_{mn}}(t).\end{array}\]
\State \quad \underline{\emph{\textbf{Step II}}}:  For each $m$, \[\;\;\!\!\begin{array}{l}
\tilde{v}^z_m(t) \!=\! \left[ {\rm var}\{z_m|y_m,z_m^s(t);v_m^s(t)\}^{-1} \!-\! ({v_m^s(t)})^{-1} \right]^{\!-1}\!\!\!\!\!,\\
\tilde{z}_{m}(t) \!=\! \tilde{v}^z_m(t) \left[\frac{{\rm E}\{z_m|z_m^s(t),v_m^s(t);y_m\}}{{\rm var}\{z_m|z_m^s(t),v_m^s(t);y_m\}} \!-\! \frac{z_m^s(t)}{v_m^s(t)}\right].\end{array}\]
\State \quad \underline{\emph{\textbf{Steps III and IV}}}:  For each $m$ and $n$, \[ \begin{array}{l}
 v^s_{mn}(t) = { |a_{mn}|^{-2}[ \tilde{v}^z_{m}(t) + v_{m}^s(t)]},\\
 x^s_{mn}(t)  = { a_{mn}^{-1}[\tilde{z}_{m}(t)- z_{m}^s(t) +a_{mn} x_{mn}^v(t)]},\\
 v_{n}^v (t)= \Big({\sum}_{m} \frac{1} {v^s_{mn}(t)} \Big)^{-1}\!\!\!\!,\; x_{n}^v(t) = v_{n}^v(t) {\sum}_{m} { \frac{x_{mn}^{s}(t)} {v^s_{mn}(t)} }. \\
\end{array}\quad\quad\]
\State \quad \underline{\emph{\textbf{Steps V and VI}}}: For each $m$ and $n$, \[\;\;\begin{array}{l}
{v}^v_{mn}(t\!+\!1) \!=\! {\rm var}\{x_n|x_n^v(t),v_n^v(t);q_n\},\\
{x}^v_{mn}(t\!+\!1) \!=\! {{\rm E}\{x_n|x_n^v(t),v_n^v(t);q_n\}} \!-\! {v}^v_{mn}(t) \frac{x_{mn}^s(t)}{{v_{mn}^s(t)}}.
\end{array}\]
\State \quad $t=t+1$
\State \textbf{While} \;{\small{$\big({\sum}_{n}|{x}^v_{n}(t) \!-\! {x}^v_{n}(t\!-\!1) | >\epsilon {\sum}_{n}|{x}^v_{n}(t) | \;{\textbf{or}}\; t\leq N_{ite}^{ese} \;\big)$}}
\State \textbf{Output:} For each $n$ and $m$,
\[  \!\!\!\!\!\!\!\!\!\!\!\!\!\!\!\!\!\!\!\!\!\!\!\!\!\!\!\!\begin{array}{l}
\hat{x}_n\!\!=\!\!{\rm E}\{x_n|x_n^v(t),\!v_n^v(t);q_n\},\\
\hat{z}_m\!\!=\!\!{\rm E}\{z_m|z_m^s(t),\!v_m^s(t);y_m\}.
 \end{array}\]
\end{algorithmic}
\end{algorithm}

\underline{\emph{\textbf{Step V (XCN $\to$ VN):}}}  
 Message update at XCN to VN for $\bf{x}_n$ uses EP with constraints $x_{n}^v(t) = x_n+ v_{n}^v(t)^{1/2}w$ and $p(x_n|q_n)$, i.e., for each $n$,
\BS\label{Eqn:XCN2VN}\begin{align}
\!\!\!\!\tilde{x}_{n}(t) &= \tilde{v}_n^x(t) \left[\frac{{\rm E}\{x_n|x_n^v(t),v_n^v(t);q_n\}}{{\rm var}\{x_n|x_n^v(t),v_n^v(t);q_n\}} - \frac{x_n^v(t)}{v_n^v(t)}\right],\\
\!\!\!\!\tilde{v}_n^x(t)&= \left[{\rm var}\{x_n|x_n^v(t),v_n^v(t);q_n\}^{-1} - {{(v_n^v(t))^{-1}}}\right]^{-1},
\end{align}\ES
where $\tilde{x}_{n}(t) = x_n+ \tilde{v}_{n}^x(t)^{1/2}w$.

\underline{\emph{\textbf{Step VI (VN $\to$ SN):}}}  
The constraints at $n$-th VN are  $\tilde{x}_{n}(t) = x_n+ \tilde{v}_{n}^x(t)^{1/2}w$ and $x^s_{mn}(t)= x_n+\tilde{v}_{mn}^s(t)^{1/2}w ,\forall m$. Message update at VN for SN are:
\BS\begin{align}
{x}^v_{mn} (t\!+\!1)&\!=\! {v}^v_{mn}(t\!+\!1) \left[\frac{\tilde{x}_{n}(t)}{\tilde{v}_n^x(t)} \!+\!\frac{x_n^v(t)}{v_n^v(t)}\!-\! \frac{x_{mn}^s(t)}{{v_{mn}^s(t)}}\right],\\
{v}^v_{mn}(t\!+\!1) &\approx [(\tilde{v}_n^x(t))^{-1} +(v_n^v(t))^{-1}  ]^{-1},\label{Eqn:VN2SN2}
\end{align}\ES
where $v_{mn}^s(t)\gg v_n^v(t)$ and $x^v_{mn}(t+1)=x_n+v^v_{mn}(t+1)^{1/2}w$.

We abandon the auxiliary variables $[\tilde{x}_{n}(t),\tilde{v}^x_{n}(t)]$, and have
\BS\label{Eqn:XCN2SN}\begin{align}
\!\!\!\!{x}^v_{mn}(t\!\!+\!\!1) &\!=\!   {{\rm E}\{x_n|x_n^v(t),v_n^v(t);q_n\}} \!-\! {v}^v_{mn}(t\!\!+\!\!1) \frac{x_{mn}^s(t)}{{v_{mn}^s(t)}},\\
\!\!\!\!\!{v}^v_{mn}(t\!\!+\!\!1) &\approx {\rm var}\{x_n|x_n^v(t),v_n^v(t);q_n\}.
\end{align}\ES
Therefore, we obtain a unified ``GMP-like" EP-MPA, and the above steps are summarized in Algorithm \ref{Alg:EP}.

\section{Equivalence between EP and GAMP/AMP}
The equivalence of EP and AMP is firstly derived in \cite{Meng2015}, based on which \cite{Meng2018SPL, Meng2018access} further proposed a unified Bayesian inference framework for the extension of AMP and VAMP to the generalized linear model. Another form of EP-based derivation for MMSE GAMP was illustrated in \cite{QZou2018}. In \cite{Zhu2019}, the max-sum GAMP was built by EP. In this section, we derive the MMSE GAMP and MMSE AMP with some approximations on the unified ``GMP-like" EP-MPA in Algorithm \ref{Alg:EP}.  
\begin{algorithm}[b!]
\caption{EP-Based MMSE GAMP}\label{Alg:GAMP}
\begin{algorithmic}[1]
\State   {\small{\textbf{Input:}   $\epsilon>0$, $N_{ite}^{ese}$}}, $\mathbf{A}$, $\bf{y}$, $\{p(x_n|q_n)\}, \{{p(y_m|z_m)}\}$.
\State \textbf{Initialization:}  $t=1$, $\{{x}^v_{n}(1)={\rm E}\{x_n|q_n\}$, ${v}^v_{n}(1)={\rm{var}}\{x_n|q_n\}, \forall n\}$, and $\{L'_m(0) = 0,\forall m\}.$
\State \textbf{Do}
\State \;\; [SN, ZCN] $\to$ [VN, XCN]: For each $m$ compute:
\[\;\;\begin{array}{l}
v^s_{m}(t)=\sum_{n}{|a_{mn}|^2{\hat{v}}^x_{n}}(t),\\
z^s_{m}(t)=   \sum_{n}{a_{mn}\hat{x}_{n}(t)}- v^s_{m}(t) L'_m(t-1),\\
L''_m(t) = { \frac{1}{v_{m}^s(t)}\left[1- \frac{{\rm var}\{z_m|y_m,z_m^s(t);v_m^s(t)\}}{v_{m}^s(t)} \right]} \\
 L'_m (t)= \frac{1}{v_{m}^s(t)}[{{\rm E}\{z_m|y_m,z_m^s(t);v_m^s(t)\}-z_{m}^s(t)}].
\end{array}\]
\State \;\; [VN, XCN] $\to$ [SN, ZCN]: For each $m$ and $n$, \[\!\!\!\!\!\!\!\!\!\!\!\!\!\!\!\!\!\!\!\!\!\!\!\!\!\!\!\begin{array}{l}
 v_{n}^v(t)= \left[\sum_{m} |a_{mk}|^{2} L''_m(t)\right]^{-1},\\
x_{n}^v(t)=\hat{x}_{n}(t) + v_{n}^v(t) \sum_{m}  a_{mn}^*L'_m(t),\\
 \hat{x}_n(t+1)={\rm E}\{x_n|x_n^v(t),v_n^v(t);q_n\},\\
 \hat{v}^x_n(t+1)={\rm var}\{x_n|x_n^v(t),v_n^v(t);q_n\}.
\end{array}\]
\State \qquad $t=t+1$
\State \textbf{While} \;{\small{$\big(  \sum_{n}|{x}^v_{n}(t) - {x}^v_{n}(t-1) | >\epsilon \sum_{n}|{x}^v_{n}(t)| \;{\textbf{or}}\; t\leq N_{ite}^{ese} \;\big)$}}
\State \textbf{Output:} For each $n$ and $m$,
\[ \!\!\!\!\!\!\!\!\!\!\!\!\!\!\!\!\!\!\!\!\!\begin{array}{l}
\hat{x}_n={\rm E}\{x_n|x_n^v(t),v_n^v(t);q_n\},\\
 \hat{z}_m={\rm E}\{z_m|z_m^s(t),v_m^s(t);y_m\}.
 \end{array}\]
\end{algorithmic}
\end{algorithm}
\subsection{Connection with GAMP}
For simplicity, we define
\BS\begin{align}
\hat{v}_n^x(t+1)&={\rm var}\{x_n|x_n^v(t),v_n^v(t);q_n\},\\
\hat{x}_n(t+1)&={\rm E}\{x_n|x_n^v(t),v_n^v(t);q_n\}.
\end{align}\ES
\begin{proposition}\label{Pro:var_approx}
For $\forall m, \forall n$, we have
\BS\begin{align}
{v}^v_{mn}(t+1)&\approx\hat{v}^x_n(t+1)\leq v_{n}^v (t)\approx\mathcal{O}(\tfrac{1}{M}){v}^s_{mn}(t).
\end{align}\ES
\end{proposition}

\begin{IEEEproof}
First, we have ${v}^v_{mn}(t+1)\approx\hat{v}^x_n(t+1)={\rm var}\{x_n|x_n^v(t),v_n^v(t);q_n\}\leq {\rm var}\{x_n|x_n^v(t),v_n^v(t)\} = v_{n}^v (t)$ since the \emph{a-priori} message $q_n$ does not increase the conditional variance. In addition, from the symmetry of the system,  $v_{n}^v (t)= \Big(\sum_{m} \frac{1} {v^s_{mn}(t)} \Big)^{-1}\approx  \mathcal{O}({M}){v^s_{mn}(t)}$. Therefore, we have ${v}^v_{mn}(t+1)=\hat{v}^x_n(t+1)\leq \mathcal{O}(\tfrac{1}{M}){v}^s_{mn}(t)$.
\end{IEEEproof}

\begin{proposition} \label{Lem:VN_approx}
Message update in \eqref{Eqn:VN2XCN} can be rewritten as
\BS\begin{align}
x_{n}^v(t)&= \hat{x}_{n}(t) + v_{n}^v(t) {\sum}_{m}  a_{mn}^*L'_m(t),\label{Eqb:VN_approxa}\\
 v_{n}^v(t) &= \Big[{\sum}_{m} |a_{mk}|^{2} L''_m(t)\Big]^{-1},\label{Eqb:VN_approxb}
\end{align}
where
\begin{align}
L'_m(t)&\equiv \frac{1}{v_{m}^s(t)}[{{\rm E}\{z_m|z_m^s(t),v_m^s(t);y_m\}-z_{m}^s(t)}],\\
L''_m(t) &\equiv { \frac{1}{v_{m}^s(t)}\Big[1-\frac{ {\rm var}\{z_m|z_m^s(t),v_m^s(t);y_m\}}{v_{m}^s(t)} \Big]}.
\end{align}\ES
\end{proposition}

\begin{IEEEproof}
See APPENDIX \ref{APP:VN_approx}.
\end{IEEEproof}

\begin{proposition} \label{Lem:SN_approx}
Message update \eqref{Eqn:VN2XCN} can be rewritten as
\BS\begin{align}
&z^s_{m}\!(t)  \! \approx\!   {\sum}_{n}{a_{mn}\hat{x}_{n}(t)}- v^s_{m}(t) L'_m(t\!\!-\!\!1),\\
&v_m^s\!(t)   \! \approx\!  {\sum}_{n} |a_{mn}|^2\hat{v}^x_{n}(t).\nonumber
\end{align}
\ES
\end{proposition}

\begin{IEEEproof}
See APPENDIX \ref{APP:SN_approx}.
\end{IEEEproof}

According to Propositions \ref{Pro:var_approx}-\ref{Lem:SN_approx}, the auxiliary variables $[{x}_{mn}^v(t),{v}^v_{mn}(t)]$ and $[{x}_{mn}^s(t),{v}^s_{mn}(t)]$ can be abandoned, and EP-MPA \ref{Alg:EP} to can be rewritten to the MMSE GAMP in Algorithm \ref{Alg:GAMP}. Therefore, we have the following lemma.

\begin{lemma}
EP-MPA is equivalent to MMSE GAMP.
\end{lemma}

For {balance systems}, we have $v^s_m(t)\!\to\! \frac{N}{M}\hat{v}_x(t)$ and $v_n^v(t)/ \hat{v}_n^x(t)\!\to\! [1\!-\langle \mathrm{var}\{z|y,z^s;v^s\}/v^s \rangle]^{-1} =[1-\langle \partial\mathrm{E}\{z|y,z^s;v^s\}/\partial z^s \rangle]^{-1}$. Therefore, the MMSE GAMP can be further simplified to
\BS\label{Eqn:GAMP_simple}\begin{align}
&\bf{z}_t = \bf{A}\hat{\bf{x}}_t-\tfrac{\hat{v}_x(t)} {\hat{v}_x(t-1)}\bf{s}_{t-1}, \;\;\bf{s}_t=\varphi(\bf{z}_t) - \bf{z}_t,\\
&\bf{x}_{t+1}  = \hat{\bf{x}}_{t}+ \tfrac{1}{1-\langle \varphi'(\bf{z}_t) \rangle} \bf{A}^H\bf{s}_t,\;\;\hat{\bf{x}}_{t+1}= \eta(\bf{x}_{t+1}),
\end{align}
\ES
where $\varphi(\bf{z}_t)={\rm E}\{\bf{z}|\bf{z}_t,\bf{y}\}$ and $\eta(\bf{x}_t)={\rm E}\{\bf{x}|\bf{x}_t,\bf{q}\}$.
{\subsection{Connection with AMP}
In AMP, from $\bf{y}=\bf{Ax}+\bf{w}$, we have
\BS\begin{align}
&\!\!\!\!{\rm var}\{z_m|z_m^s(t),v_m^s(t);y_m\} = \big[\sigma_n^{-2} + (v_m^s)^{-1}\big]^{-1},\nonumber\\
&\!\!\!\!{\rm E}\{z_m|z_m^s(t),v_m^s(t);y_m\} \!\!=\!\! \big[\sigma_n^{-2} \!\!+\!\! {v_m^s}^{\!\!\!-1}\big]^{\!-1}[\sigma_n^{\!-2}y_m \!+ \!{v_m^s}^{\!\!\!-1}\! z_m^s].\nonumber
\end{align}\ES
Thus,
\begin{align}\label{Eqn:AMP_approx}
L''_m(t) = (\sigma_n^{2} +v_m^s(t))^{-1}, \;\;
L'_m(t)  = \frac{y_m-z_m^s(t)}{\sigma_n^{2} +v_m^s(t)}.
\end{align}

From \eqref{Eqb:VN_approxa} and \eqref{Eqn:AMP_approx}, we have
\BE
x^v_n(t)=\hat{x}_n(t) +  {\sum}_{m}a_{mn}^*[y_m-z_m^s(t)].\label{Eqn:var_average4}
\EE
Then, we have the following lemma.

\begin{lemma}\label{Lem:AMP}
EP-MPA can be rewritten to AMP.
\end{lemma}

\begin{IEEEproof}
See APPENDIX \ref{APP:AMP_lem}.
\end{IEEEproof}
}

{\section{Numerical Results}
We study a clipped compressed sensing problem where $\bf{x}$ follows a symbol-wise Bernoulli-Gaussian distribution, i.e. $\forall i$,
\BE
{x_i}\sim\left\{ \begin{array}{l}
0, \qquad\qquad\;\;\; \mathrm{probability} = 1-\lambda,\\
\mathcal{N}(0,{\lambda ^{ - 1}}),\quad \mathrm{probability} = \lambda,
\end{array} \right.
\EE
where  the variance of $x_i$ is normalized to 1. In addition, $\bf{y}$ is a non-linear clipping noisy function of $
\bf{z}$, i.e.
\BE
\bf{y} = Q(\bf{z}) + \bf{n},
\EE
where $\bf{n}\sim \mathcal{N}(0, \sigma^2\bf{I})$ is a Gaussian noise vector. Let $\theta$ be a positive threshold, $Q(\cdot)$ is a symbol-wise function given by
\BE
  Q(z)= \left\{ \begin{array}{l}
  \!\!\!\!-\theta, \quad   z\leq \theta\\
z,  \quad   -\theta < z <\theta\\
\theta, \quad   z\geq \theta
\end{array} \right..
\EE
The transmit \emph{signal-to-noise-ratio} (SNR) is defined as $SNR = \mr{E}\{\|x_i\|^2\}/\mr{E}\{\|n_j\|^2\}=\sigma^{-2}$.

\begin{figure}[t]
  \centering
  \includegraphics[width=9cm]{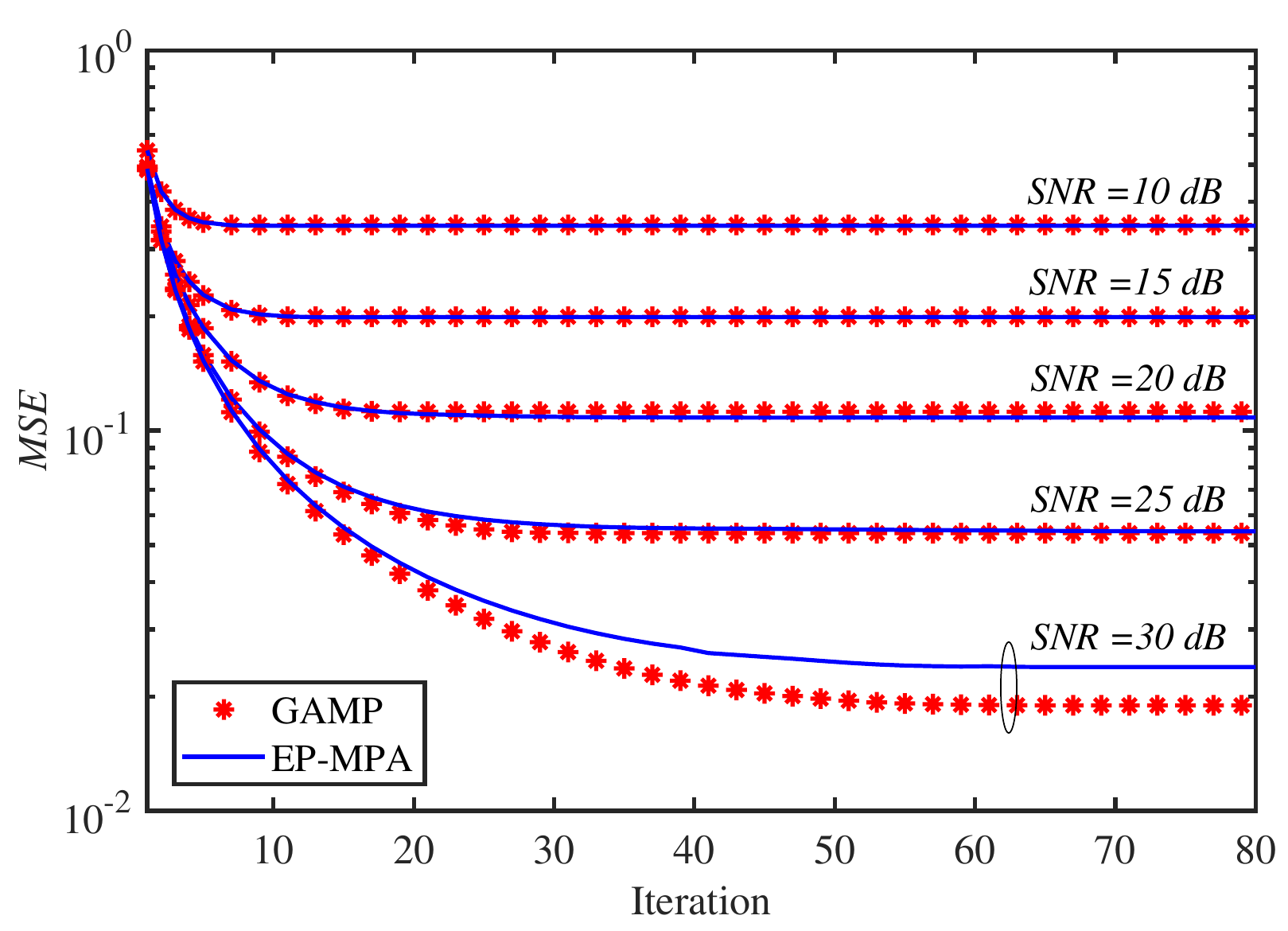}\\\vspace{-0.1cm}
  \caption{MSE comparison between EP-MPA and GAMP for clipped compressed sensing, where $M=N=10^4$, $\lambda=0.5$, $\theta=1$, $SNR=\{10, 15, 20, 25, 30\}$ (dB). }\label{Fig:MSE_Com}
\end{figure}
Fig. \ref{Fig:MSE_Com} shows the mean square error (MSE) comparison between the original EP-MPA in Algorithm \ref{Alg:EP} and the GAMP in \eqref{Eqn:GAMP_simple}. The simulation results show that the MSE curves of EP-MPA and GAMP are well-matched, which verifies the equivalence of EP-MPA and GAMP. Note that this equivalence is based on the assumption of $N\to \infty$. In high SNR, it is rational that EP-MPA is slightly worse than GAMP for finite $N$. In addition, the variance updates are averaged in \eqref{Eqn:GAMP_simple}, which also leads to the difference between the original EP-MPA\footnote{The variance of EP-MPA in Algorithm \ref{Alg:EP} may be negative, which should be positive. This leads to the performance loss of EP-MPA. Some modifications can be used to avoid the negative variance. For more details, refer to \cite{Cespedes2014}.} and the GAMP in \eqref{Eqn:GAMP_simple}.}

\section{Conclusion}
In this correspondence, an EP-MPA is considered for the general measurement channel. We prove that EP-MPA is equivalent to the well known GAMP and AMP by the omission of high-order terms, which are negligible in large system limit. Since the proposed EP-MPA is constructed with a unified ``GMP-like" message passing rule, which is  easier to understand than the derivation of GAMP and AMP, these results results offer a new insight into GAMP and AMP, and provide hints to solving more general non-linear problems.

\appendices
\section{Proof of Proposition \ref{Lem:VN_approx}}\label{APP:VN_approx}
First, we prove \eqref{Eqb:VN_approxa}.
\BS\begin{align}
   &\frac{1} {|a_{mn}|^{2}v^s_{mn}(t)}
\approx [{\tilde{v}^z_{m}(t) + v_{m}^s(t)}]^{-1}\label{Eqn:L_m2}\\
 &= { \frac{1}{v_{m}^s(t)}\left[1\!-\!\frac{ {\rm var}\{z_m|z_m^s(t),v_m^s(t);y_m\}}{v_{m}^s(t)}  \right]} =L''_m(t),\label{Eqn:L_m3}
\end{align}\ES
where \eqref{Eqn:L_m2}  follows \eqref{Eqn:SN2VN}, and \eqref{Eqn:L_m3} follows \eqref{Eqn:ZCN2SN}. Hence,
\BE
 v_{n}^v(t) = \Big[{\sum}_{m} \frac{1}{v^s_{mn}(t)} \Big]^{-1} = \Big[{\sum}_{m} |a_{mn}|^{2} L''_m(t)\Big]^{-1}.
\EE

Then, we prove \eqref{Eqb:VN_approxb}.
\BS\begin{align}
 &\frac{  \tilde{z}_{m}(t)- z_{m}^s(t) } {|a_{mn}|^2v^s_{mn}(t) }
 =\frac{  \tilde{z}_{m}(t)- z_{m}^s(t)} {{\tilde{v}^z_{m}(t) + v_{m}^s(t)}}\label{L1_m1}\\
 &= \frac{1}{v_{m}^s(t)}[{{\rm E}\{z_m|z_m^s(t),v_m^s(t);y_m\}\!-\!z_{m}^s(t)}] = L'_m(t),\label{L1_m2}
\end{align}\ES
where \eqref{L1_m1} follows \eqref{Eqn:L_m2}, and \eqref{L1_m2} is from \eqref{Eqn:ZCN2SN}.
Then,
\BS\label{Eqn:u_v_approx}\begin{align}
\!\!&x_{n}^v(t) 
= v_{n}^v(t) {\sum}_{m}  \Big[ \frac{x_{mn}^v(t)}  { v^s_{mn}(t)} +  a_{mn} ^* \frac{\tilde{z}_{m}(t)- z_{m}^s(t)}  {|a_{mn}|^2v^s_{mn}(t)}\Big] \nonumber\\
\!\!\!\!\!  &\approx \!v_{n}^v \!(t)\!\!\sum_{m} \!\! \Big[ \!  \tfrac{{\rm E}\{x_n|x_n^v(t\!-\!1),v_n^v(t\!-\!1);q_n\!\}\!-\! v_{mn}^v\!(t)\tfrac{x_{mn}^s\!(t\!-\!1)}{v_{mn}^s\!(t\!-\!1)}}  { v^s_{mn}(t)}  \!+\!   a_{mn}^*L'_m(t)\!\Big]  \nonumber\\
\!\!\!\!\!  &= \hat{x}_n(t) + v_{n}^v(t) {\sum}_{m}  a_{mn}^*L'_m(t),
\end{align}\ES
where $ v_{n}^v (t){\sum}_{m}   \tfrac{  v_{mn}^v(t)\tfrac{x_{mn}^s(t-1)}{v_{mn}^s(t-1)}}  { v^s_{mn}(t)} \leq O(\tfrac{1}{M})x_{mn}^s(t-1)$ is negligible since $\tfrac{{v}^v_{n}(t)}{{v}^s_{mn}(t)}  \approx \mathcal{O}(\tfrac{1}{M})$ and $\tfrac{{v}^v_{mn}(t)}{{v}^s_{mn}(t-1)}  \leq \mathcal{O}(\tfrac{1}{M})$.

\section{Proof of Proposition \ref{Lem:SN_approx}}\label{APP:SN_approx}
From Proposition \ref{Pro:var_approx}, we have $v_m^s(t) \approx {\sum}_{n} |a_{mn}|^2\hat{v}_{n}^x(t)$, and
\begin{align}
\!\!\!\!\!\!&z^s_{m}(t)
=  {\sum}_{n}{a_{mn}\Big[\hat{x}_n(t) - {v}^v_{mn}(t)\frac{x_{mn}^s(t-1)}{{v_{mn}^s(t-1)}}\Big] } \nonumber\\
\!\!\!\!\!\!\!\!\!\!&= \!\! {\sum}_{n}\!a_{mn}\Big[{\hat{x}_{n}\!(t) \!-\!  v_{mn}^v\!(t)\Big[a_{mn}^HL'_m(t\!\!-\!\!1) \!+\! \frac{x_{mn}^v(t\!\!-\!\!1)}{{v_{mn}^s(t\!\!-\!\!1)}}\Big] }\Big] \nonumber\\
\!\!\!\!\!\!\!\!\!\!&={\sum}_{n}{a_{mn}\hat{x}_{n}}(t)- L'_m(t-1)  {{\sum}_{n}{|a_{mn}|^2{{v_{mn}^v(t)}}}} \nonumber\\
\!\!\!\!\!\!\!\!\!\!&={\sum}_{n}{a_{mn}\hat{x}_{n}}(t)- v^s_{m} (t) L'_m(t-1),\label{Eqn:pro3_eq4}
\end{align}
where the first two equations are from \eqref{Eqn:XCN2SN} and \eqref{Eqn:u_v_approx}, and the third from ${\sum}_{n} a_{mn} {v}^v_{mn}(t) \frac{u_{mn}^v(t-1)}{{v_{mn}^s(t-1)}}$  is negligible since $a_{mn}\sim \mathcal{CN}(0,1/M)$ and ${v}^v_{mn}(t)/{v}^s_{mn}(t-1)\leq \mathcal{O}(\tfrac{1}{M})$.

\section{Proof of Lemma \ref{Lem:AMP}}\label{APP:AMP_lem}
According to the i.i.d. property, we have
\BS\begin{align}
\hat{v}_n^x(t) &\approx \tfrac{1}{N}{\sum}_{n}{\rm var}\{x_n|x^v_n(t),v^v_n(t);q_n\}= \hat{v}^x(t),\label{Eqn:var_average1}\\
v_m^s(t) &= {\sum}_{n} |a_{mn}|^2\hat{v}_n^x(t) \approx \tfrac{N}{M}\hat{v}^x(t)=v^s(t),\label{Eqn:var_average2}\\
v_n^v(t) &= \Big[{\sum}_{m} |a_{mn}|^2L''_m(t)\Big]^{-1} \!\!\!\!\approx \!\sigma_n^2 + v^s(t)=v^v(t),\label{Eqn:var_average3}
\end{align}\ES
where \eqref{Eqn:var_average2} is from ${\sum}_{n} |a_{mn}|^2\to N/M$, and \eqref{Eqn:var_average2} from \eqref{Eqn:AMP_approx}.

Let $\bf{x}^v_t=[{x}^v_1(t),\cdots,{x}^v_N(t)]$, $\hat{\bf{x}}_t=[\hat{{x}}_1(t),\cdots,{\hat{x}}^v_N(t)]$, ${\bf{L}}'_t=[L'_1(t),\cdots,L'_M(t)]$, and ${\bf{z}}_t=[z_1^s(t),\cdots,z^s_M(t)]$. From \eqref{Eqn:var_average4} and \eqref{Eqn:AMP_approx}, we have
\BS\label{Eqn:lem_amp}\begin{align}
&\bf{x}^v_t   = \hat{\bf{x}}_t+ \bf{A}^H(\bf{y}-\bf{A}\hat{\bf{x}}_t + v^s(t) {\bf{L}}'_{t-1}),\label{Eqn:lem_amp0}\\
&\hat{\bf{x}}_{t+1}  = \eta_t({\bf{x}^v_t})={\rm E}\{\bf{x} |\bf{x}^v(t),\bf{v}^v(t);\bf{q}\}.
\end{align}
where
\BE
v^s(t) {\bf{L}}'_{t-1} = \frac{v^s(t)}{ v^v(t-1)} (\bf{y}-\bf{z}^s_{t-1}),
\EE
and
\begin{align}
&\frac{v^s(t)}{v^v(t-1)}   = \frac{{\sum}_{n}{|a_{mn}|^2{v}_{mn}^v}(t) }{v^v(t-1)}
\approx \frac{\frac{1}{M} {\sum}_{n} {{\hat{v}}_n^x }(t) }{v^v(t-1)} \label{Eqn:lem_amp1}\\
& = \frac{\frac{1}{M}{\sum}_{n} { {v^v_n(t-1)\frac{\partial {\rm E}\{x_k|x_n^v(t-1),v_{n}^v(t-1);q_n\}}{\partial x_{n}^v({t-1})}} }}{v^v({t-1}) }\label{Eqn:lem_amp2}\\
 & = \tfrac{1}{M}{\sum}_{n} \eta'_{t-1}(x_n^v(t-1))  =\tfrac{N}{M}\langle\eta'_{t-1}(\bf{x}^v_{t-1})\rangle,\label{Eqn:lem_amp3}
\end{align}\ES
where \eqref{Eqn:lem_amp1} is due to $ |a_{mn}|^2\approx \tfrac{1}{M}$, \eqref{Eqn:lem_amp2} follows ${{\hat{v}}_n^x }(t) = { {v^v_n(t-1)\frac{\partial {\rm E}\{x_k|x_n^v(t-1),v_{n}^v(t-1);q_n\}}{\partial x_{n}^v({t-1})}} }$, and \eqref{Eqn:lem_amp3} is due to $\langle\eta'_{t-1}(\bf{x}^v_{t-1})\rangle\equiv\tfrac{1}{N}{\sum}_{n} \eta'_{t-1}(x_n^v(t-1))$.

With \eqref{Eqn:lem_amp} and $\bf{z}_t=\bf{A}^H(\bf{x}^v_{t}-\hat{\bf{x}}_{t})$, we obtain AMP below.
\BS\label{Eqn:AMP_alg}\begin{align}
&\bf{z}_t  = \bf{y}-\bf{A}\hat{\bf{x}}_t +  \tfrac{N}{M}\langle\eta'_{t-1}(\bf{A}^H\bf{z}_{t-1}+\hat{\bf{x}}_{t-1})\rangle\bf{z}_{t-1},\\
&\hat{\bf{x}}_{t+1}  = \eta_t(\hat{\bf{x}}_t+\bf{A}^H\bf{z}_{t}).
\end{align}\ES


\end{document}